# Population-Level Decision Curve Analysis May Mislead the Evaluation of Prediction Model Usefulness under Subgroup Utility Heterogeneity

**Junfeng Wang (1), Kim Zhipei Wang (1), Ben Van Calster (2,4,5), Nan van Geloven (3), Ewout Steyerberg (4), Laure Wynants (1,2,5)**

*1. Department of Epidemiology, CAPHRI Care and Public Health Research Institute, Maastricht University, Maastricht, The Netherlands*
*2. Department of Development and Regeneration, KU Leuven, Leuven, Belgium*
*3. Department of Biomedical Data Sciences, Leiden University Medical Center, Leiden, the Netherlands*
*4. Julius Center for Health Sciences and Primary Care, University Medical Center Utrecht, Utrecht University, Utrecht, The Netherlands*
*5. Leuven Unit for Health Technology Assessment Research (LUHTAR), KU Leuven, Leuven, Belgium*

# Abstract

## Background

Decision curve analysis (DCA) is widely used to evaluate the clinical usefulness of prediction models using net benefit (NB). In practice, population-level NB is often interpreted as a proxy for population-level expected utility, but the assumptions underlying this interpretation are rarely examined. We aimed to formally characterize when population-level DCA may lead to incorrect conclusions about prediction model usefulness in the presence of subgroup utility heterogeneity.

## Methods

We considered comparisons between a prediction-driven treatment strategy and default treatment strategies in a population comprising subgroups with potentially different utility values. We analytically examined the relationship between population-level ΔUtility and ΔNB and derived the "inconsistency region", defined as combinations of subgroup-specific ΔNB values for which population-level NB and utility favor different strategies. We also developed a practical framework to assess the robustness of population-level DCA conclusions in the presence of subgroup utility heterogeneity.

## Results

We demonstrate that conflicting subgroup-specific ΔNB results provide an empirical warning signal that population-level NB may favor prediction model use even when population utility favors the comparator strategy. The inconsistency region becomes larger when subgroup sizes are more similar and subgroup-specific $a - c$ values are more dissimilar, where $a - c$ represents the incremental utility of a true positive relative to a false negative. When $a - c$ varies across subgroups, population-level NB aggregates quantities expressed on different implicit utility scales and may therefore yield conclusions that conflict with population utility. A real-world case study illustrates the practical relevance of this problem.

## Conclusions

Interpreting population-level NB as a proxy for population utility implicitly assumes homogeneous $a - c$ values across subgroups. When this assumption is uncertain, subgroup-specific DCA can identify potential inconsistencies, and our framework can assess whether subgroup utility heterogeneity may invalidate conclusions based on population-level NB.

## Highlights

• Interpreting population-level net benefit (NB) as a proxy for population utility implicitly assumes that the value of a true positive relative to a false negative is homogeneous across subgroups.

• When this assumption is violated, population-level NB may favor prediction model use while population-level utility favors the comparator strategy, or vice versa, particularly when subgroups are similar in size and differ substantially in this value.

• We propose a practical framework to assess whether subgroup utility heterogeneity may invalidate conclusions based on population-level NB.

# Background

Decision curve analysis (DCA) is widely used and recommended for evaluating the clinical usefulness of diagnostic tests and risk prediction models when positive predictions trigger an intervention, such as an invasive diagnostic procedure, referral, or treatment initiation [1-3]. Although DCA does not require explicit utility elicitation, it is fundamentally utility-based because net benefit (NB) is derived from an implicit utility model in which the threshold probability encodes the relative utilities of different decision outcomes [2]. Specifically, DCA builds on the relationship that the relative value of a false positive versus a true negative, compared with that of a true positive versus a false negative, equals $\frac{t^*}{1-t^*}$, where the threshold probability $t^*$ represents the outcome risk above which one would be willing to intervene. Within this framework, NB is defined as the proportion of true positives minus the proportion of false positives weighted by $\frac{t^*}{1-t^*}$ [2]. A key advantage of DCA is that NB can be readily calculated during model validation as a surrogate for clinical utility. Assessing NB across a range of threshold probabilities can therefore provide insights into prediction model usefulness beyond traditional measures such as sensitivity, specificity, and the area under the receiver operating characteristic (ROC) curve.

Most applications evaluate NB across the entire study population, implicitly adopting a "one-size-fits-all" perspective. Previous work has highlighted several reasons to consider subgroups in DCA. Kerr et al. [3] noted that risk thresholds, distributions of predicted event probabilities, and outcome prevalence may differ across subgroups and advised against deriving subgroup-specific decisions by reading NB at subgroup-specific thresholds from a decision curve based on the combined population [4]. More recently, Benitez-Aurioles et al. [5] highlighted differences in outcome prevalence across subgroups and introduced a modified subgroup net benefit (sNB) from a fairness perspective, which requires more detailed information on utility values than conventional NB. Conversely, Vickers et al. [6] argued that evidence supporting distinct DCA conclusions across subgroups should be compelling and, given the limited data often available within subgroups, recommended focusing first on population-level DCA when evaluating a model for clinical use. This approach implicitly relies on population-level NB providing a valid representation of population utility when the population comprises heterogeneous subgroups.

In this paper, we show that this correspondence can break down: population-level NB may favor one strategy while population-level utility favors another when subgroup-specific utilities differ. We focus on conventional DCA for diagnostic and prognostic prediction models, rather than DCA for models that predict differences in outcome risk under alternative treatment strategies [7, 8]. We first present a motivating example in which NB and utility diverge even under restrictive conditions, then formally characterize the conditions under which such inconsistencies arise and examine them through scenario analyses. Finally, we propose a practical framework for assessing the robustness of population-level DCA in the presence of subgroup utility heterogeneity and illustrate its application in a real-world case study.

# A motivating example

Consider a diagnostic prediction model for disease presence that triggers an intervention when the predicted probability exceeds a threshold $t$. Let $a$, $b$, $c$, and $d$ denote the utilities of true positives, false positives, false negatives, and true negatives, respectively (Table 1). The relationship between the optimal risk threshold $t^*$ and these utilities has been established elsewhere [2] and is given in the formula below. For this example, assume a population comprising two subgroups with identical optimal thresholds:

$$t_1^* = t_2^*$$

$$\frac{1-t_1^*}{t_1^*} = \frac{a_1 - c_1}{d_1 - b_1} = \frac{1-t_2^*}{t_2^*} = \frac{a_2 - c_2}{d_2 - b_2}$$

In this example, we assume that $c$ and $d$ are the same across subgroups but allow $a$ and $b$ to differ. To make the example concrete, utilities are expressed in terms of quality of life (QoL). In both groups, QoL is 100 in the absence of disease or intervention-related complications ($d_1 = d_2 = 100$) and 34 when disease is missed and no intervention is given ($c_1 = c_2 = 34$). Intervention after detection increases QoL to 94 in group 1 ($a_1 = 94$) but to only 79 in group 2 ($a_2 = 79$). Unnecessary intervention is also less harmful in group 2: QoL among false positives is 96 in group 1 ($b_1 = 96$) and 97 in group 2 ($b_2 = 97$). Thus, the incremental utility of a true positive relative to a false negative ($a - c$) is smaller in group 2, as is the harm of a false positive relative to a true negative ($d - b$). However, because the ratio $(d - b)/(a - c)$ is identical in both groups, they have the same optimal threshold. (Thus, even with identical outcome prevalence and optimal thresholds across subgroups, population-level NB and utility can rank the same strategies differently. Figure 1 shows ΔNB and ΔUtility for the model relative to the best default strategy in each subgroup and in the overall population, where $\Delta NB = NB_{\text{model}} - \max(NB_{\text{treat none}}, NB_{\text{treat all}})$ and $\Delta Utility = U_{\text{model}} - \max(U_{\text{treat none}}, U_{\text{treat all}})$. This example illustrates that when the utility scale differs across subgroups, maximizing population-level NB does not necessarily maximize population utility.

Table 1)

Each subgroup comprises 100,000 individuals with a disease prevalence of 50%. Specificity is 98% in both groups, whereas sensitivity is 90% in group 1 and 98% in group 2 (Table 2). Given identical thresholds and disease prevalence across subgroups, a DCA comparing the model's NB with the default strategies "treat all" and "treat none" provides a reasonable starting point [6]. With equal subgroup sizes, the population-level NB of the model is 0.4693, exceeding that of both "treat all" (NB=0.4667) and "treat none" (NB=0). Population-level NB therefore favors the model.

The conclusion reverses when the strategies are compared on the utility scale. Population utility can be calculated directly by multiplying the utility assigned to each classification outcome ($a$ to $d$) by the corresponding number of individuals and aggregating across the two subgroups:

$$U_{1+2}(t^*) = \frac{1}{N}\left(n_1 U_1(t^*) + n_2 U_2(t^*)\right) = \frac{1}{N}\left(a_1 TP_1(t^*) + b_1 FP_1(t^*) + c_1 FN_1(t^*) + d_1 TN_1(t^*) + a_2 TP_2(t^*) + b_2 FP_2(t^*) + c_2 FN_2(t^*) + d_2 TN_2(t^*)\right)$$

Using this calculation, "treat all" yields a population utility of 183, slightly higher than that of the model (182.98), and is therefore favored on the utility scale (Figure 1). Thus, even with identical outcome prevalence and optimal thresholds across subgroups, population-level NB and utility can rank the same strategies differently. Figure 1 shows ΔNB and ΔUtility for the model relative to the best default strategy in each subgroup and in the overall population, where $\Delta NB = NB_{\text{model}} - \max(NB_{\text{treat none}}, NB_{\text{treat all}})$ and $\Delta Utility = U_{\text{model}} - \max(U_{\text{treat none}}, U_{\text{treat all}})$. This example illustrates that when the utility scale differs across subgroups, maximizing population-level NB does not necessarily maximize population utility.

*Table 1 Utilities as quality of life after diagnosis & intervention*

| Group 1 – larger intervention benefit | Has the illness | Does not have the illness | a-c | d-b | (1-t)/t | t |
|---|---|---|---|---|---|---|
| P>t (predicted positive) | $a_1$=94 | $b_1$=96 | 60 | 4 | 15 | 0.0625 |
| P<t (predicted negative) | $c_1$=34 | $d_1$=100 | | | | |
| | | | | | | |
| Group 2 – smaller intervention benefit | Has the illness | Does not have the illness | | | | |
| P>t (predicted positive) | $a_2$=79 | $b_2$=97 | 45 | 3 | 15 | 0.0625 |
| P<t (predicted negative) | $c_2$=34 | $d_2$=100 | | | | |

*Table 2 Confusion matrices for classification performance evaluation (Numbers in thousands)*

| GROUP 1 PERFORMANCE | Has the illness | Does not have the illness |
|---|---|---|
| P>t (predicted positive) | 45 | 1 |
| P<t (predicted negative) | 5 | 49 |
| Total | 50 | 50 |
| | | |
| GROUP 2 PERFORMANCE | Has the illness | Does not have the illness |
| P>t (predicted positive) | 49 | 1 |
| P<t (predicted negative) | 1 | 49 |
| Total | 50 | 50 |
| | | |
| GROUP 1+2 PERFORMANCE | Has the illness | Does not have the illness |
| P>t (predicted positive) | 94 | 2 |
| P<t (predicted negative) | 6 | 98 |
| Total | 100 | 100 |

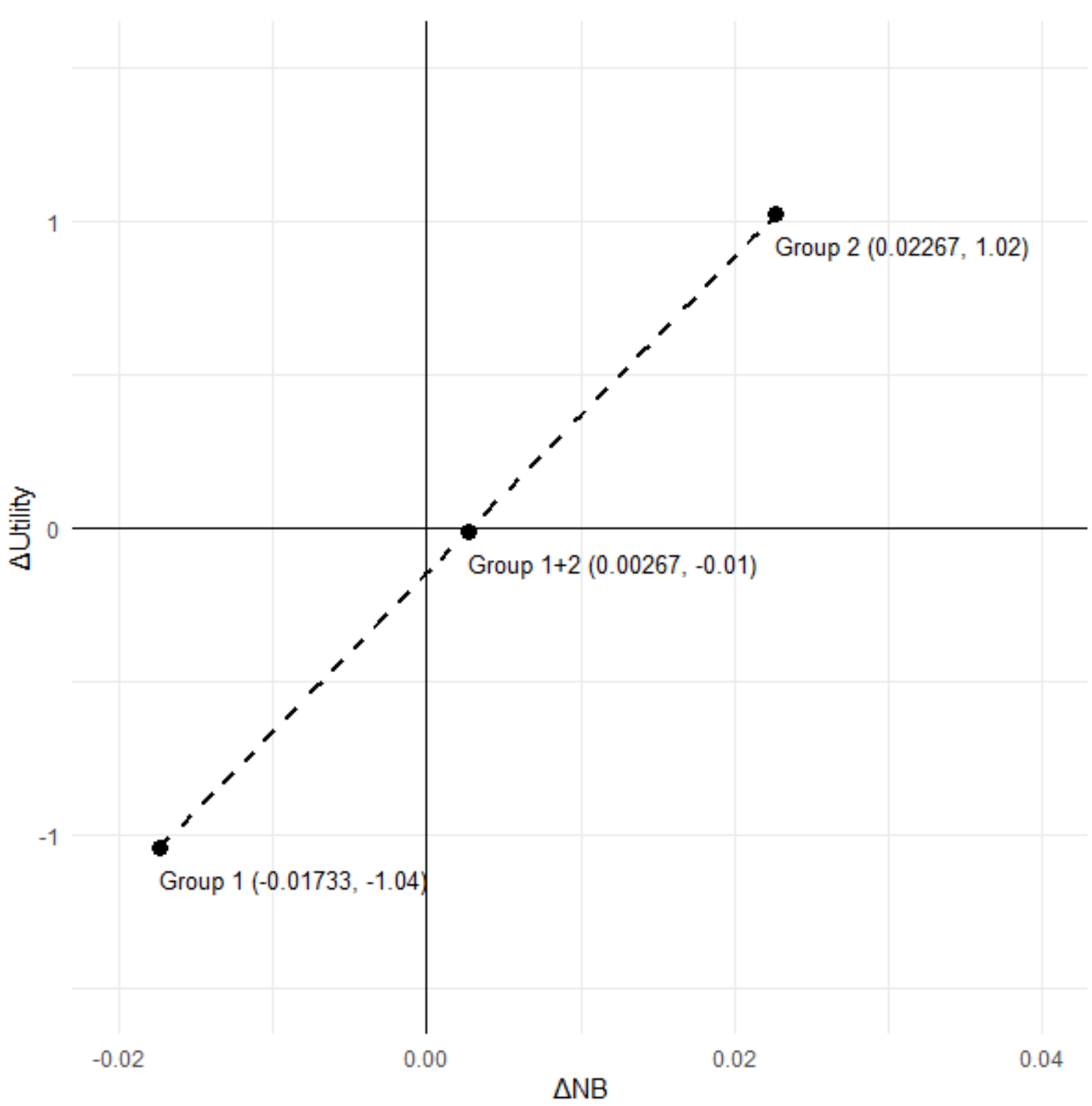

*Figure 1 Relation between ΔNB ($NB_{model}$ - max($NB_{treat\ none}$, $NB_{treat\ all}$)) and ΔUtility ($U_{model}$ - max($U_{treat\ none}$, $U_{treat\ all}$)) in subgroups and the population*

*Note: the dotted line represents all possible weighted averages of subgroup-specific ΔNB/ΔUtility, while the dots represent the group-specific and sample-size-weighted ΔNB/ΔUtility.*

# Origin of inconsistency

## Inconsistency due to weighting

To understand the origin of the inconsistency illustrated above, we now consider the mathematical relationship between utility and NB. The key issue arises when subgroup-specific quantities expressed on different utility scales are aggregated at the population level.

We start with the expected utility of a model-based strategy:

$$U(t) = \frac{1}{N}(aTP(t) + bFP(t) + cFN(t) + dTN(t))$$

*Equation 1*

For given values of $a$, $b$, $c$, and $d$, the optimal threshold $t^*$ that maximizes expected utility is determined by [2]:

$$\frac{1-t^*}{t^*} = \frac{a-c}{d-b}$$

Substituting this relationship into Equation 1 yields:

$$U(t^*) = \frac{a-c}{N}\left(TP(t^*) - \frac{t^*}{1-t^*}FP(t^*)\right) + (c-d)\pi + d$$

*Equation 2*

where $\pi$ denotes outcome prevalence (event rate). Terms that do not depend on the model, $(c-d)\pi + d$, can either be omitted when comparing strategies or removed by expressing utility relative to the “treat none” strategy [9]. The remaining quantity leads to the standard definition of NB, which expresses utility in units of $a-c$ [2]:

$$NB(t^*) = \frac{1}{N}\left(TP(t^*) - \frac{t^*}{1-t^*}FP(t^*)\right)$$

*Equation 3*

The utility equation can therefore be written as:

$$U(t^*) = (a-c)NB(t^*) + (c-d)\pi + d$$

*Equation 4*

Without loss of generality, we illustrate the relationship using a comparison between the model and “treat all”. The corresponding derivations for comparisons of the model with “treat none”, “treat all” with “treat none”, and two competing models are provided in the Appendix. From Equation 4:

$$\begin{aligned}\Delta U = U(t^*) - U(Tx\ all) &= (a-c)NB(t^*) + (c-d)\pi + d - \big((a-c)NB(Tx\ all) + (c-d)\pi + d\big)\\ &= (a-c)NB(t^*) - (a-c)NB(Tx\ all) = (a-c)\Delta NB\end{aligned}$$

*Equation 5*

Thus, within a homogeneous population, the difference in utility between two strategies is proportional to their difference in NB, with $a-c$ as the scaling factor. Provided that $a > c$, ΔNB and ΔUtility therefore always favor the same strategy.

Now consider a population comprising two subgroups with subgroup-specific values $a_1 - c_1$ and $a_2 - c_2$ and sample sizes $n_1$ and $n_2$. For a common threshold probability across subgroups, population-level ΔNB is the sample-size-weighted average of subgroup-specific ΔNB (a formal proof is provided in the Appendix):

$$\Delta NB = \Delta NB_{1+2} = \frac{1}{N}(n_1\Delta NB_1 + n_2\Delta NB_2)$$

By contrast, because $\Delta U_i = (a_i - c_i)\Delta NB_i$, population-level ΔUtility is proportional to a weighted combination of subgroup-specific ΔNB, with weights $n_1(a_1 - c_1)$ and $n_2(a_2 - c_2)$:

$$\Delta U_{1+2} = \frac{1}{N}(n_1\Delta U_1 + n_2\Delta U_2) = \frac{1}{N}(n_1(a_1 - c_1)\Delta NB_1 + n_2(a_2 - c_2)\Delta NB_2)$$

Population-level ΔNB therefore weights subgroup-specific ΔNB only by subgroup size, whereas population-level ΔUtility additionally weights each subgroup by $a_i - c_i$. When $a_1 - c_1 \neq a_2 - c_2$, the two population-level measures can consequently favor different strategies: $\Delta NB_{1+2}>0$ while $\Delta U_{1+2}<0$, or vice versa. In other words, subgroup-specific NB values are expressed on different implicit utility scales when $a - c$ differs across subgroups, and aggregating them by subgroup size alone need not preserve the population-utility ranking of strategies.

## Analytic characterization of the inconsistency region

The preceding derivation shows that population-level NB may not always provide a reliable proxy for population utility when $a - c$ differs across subgroups. We next formally characterize the conditions under which population-level ΔNB and ΔUtility favor different strategies, that is, $\Delta U_{1+2}\times\Delta NB_{1+2}<0$.

For a more concise derivation, let $a_1 - c_1 = k_1$ and $a_2 - c_2 = k_2$, and denote the subgroup-specific differences in NB as $\Delta NB_1 = x$ and $\Delta NB_2 = y$.

$$\Delta U_{1+2} \times \Delta NB_{1+2} = \frac{1}{N}(n_1k_1x + n_2k_2y) \times \frac{1}{N}(n_1x + n_2y)$$
$$= \frac{1}{N^2}({n_1}^2k_1x^2 + n_1n_2(k_1 + k_2)xy + {n_2}^2k_2y^2) = \frac{1}{N^2}f(x, y)$$

When $k_1 \neq k_2$, $f(x, y)$ is an indefinite quadratic form (proof provided in the Appendix). Consequently, there exist combinations of $(x, y)$ for which $f(x, y) < 0$. We define these combinations as the inconsistency region: the set of subgroup-specific ΔNB values for which population-level ΔNB and population-level ΔUtility favor different strategies.

The boundaries of the inconsistency region are given by $f(x, y) = 0$, which consists of two straight lines intersecting at the origin, with slopes $s_1 = -\frac{n_1}{n_2}$ and $s_2 = -\frac{n_1k_1}{n_2k_2}$ (proof provided in the Appendix) (Figure 2). When $(\Delta NB_1, \Delta NB_2)$ falls within the shaded region in Figure 2, population-level NB and utility favor different strategies: NB favors prediction model use whereas utility favors the comparator strategy, or vice versa.

If $a_1 - c_1 = a_2 - c_2$, the two weighting schemes coincide and no inconsistency region exists; population-level ΔNB and ΔUtility then always favor the same strategy. To examine how the inconsistency region behaves under practically relevant combinations of thresholds, model performance, and subgroup characteristics, we additionally conduct scenario analyses in the Appendix.

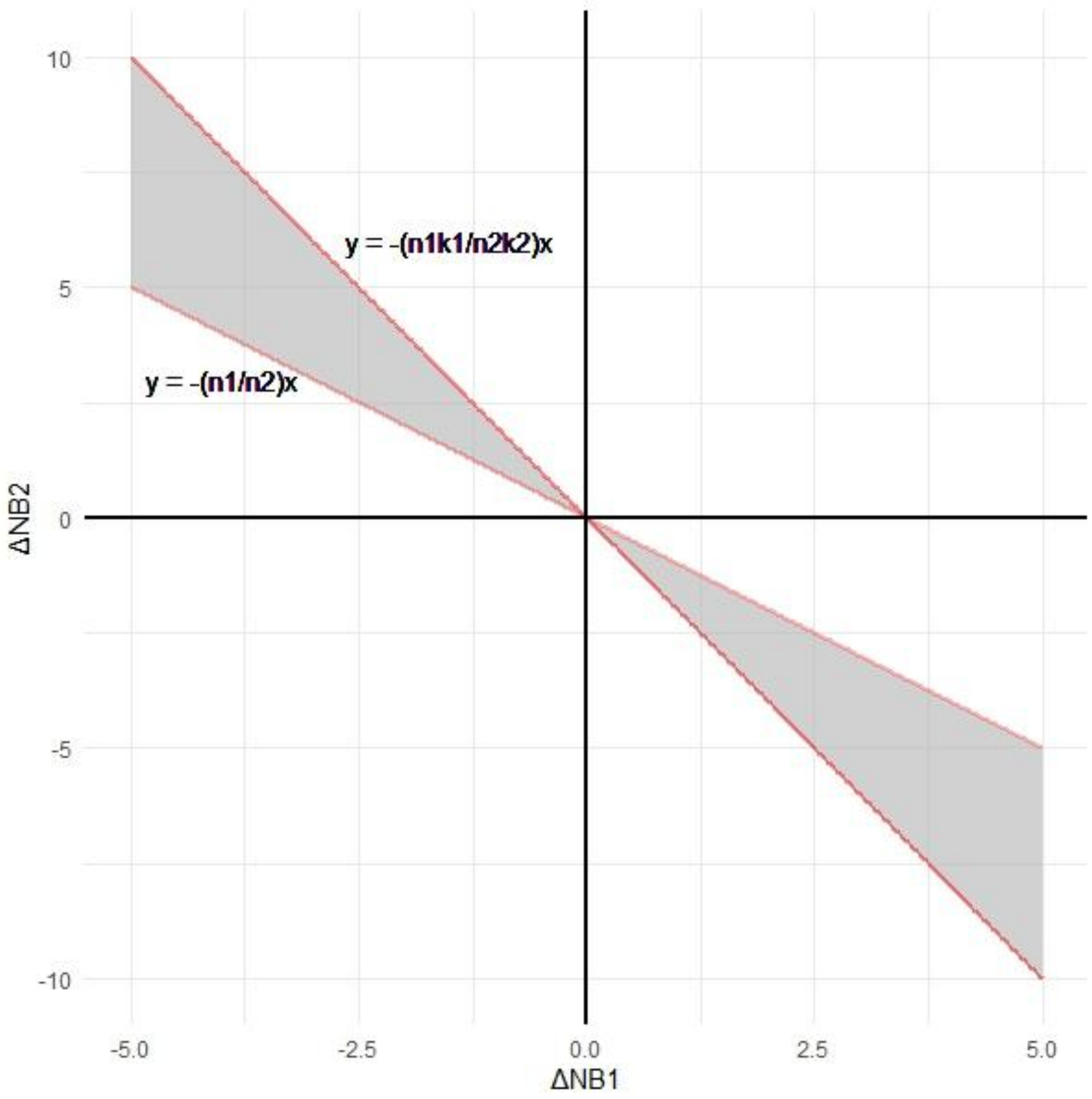

*Figure 2 Inconsistency region determined by the subgroup size ratio $n_1/n_2$and subgroup-specific incremental utilities of a true positive relative to a false negative ($a_1 - c_1 = k_1$ and $a_2 - c_2 = k_2$). ΔNB denotes the difference in net benefit between the model and a comparator strategy, such as "treat all", "treat none", or another model.*

## Utility is the gold standard, while NB is not

We use the term "gold standard" to denote the quantity that decision-making ultimately seeks to optimize. In this sense, utility is the reference quantity, whereas NB is a convenient surrogate whose correspondence with utility depends on the underlying utility scale.

In a guide to interpreting DCA, Vickers et al. [1] used the example of a wine importer to illustrate NB as analogous to net profit. True positives (TPs) are analogous to income in US dollars and false positives (FPs) to expenditure in euros, with $(d-b)/(a-c)$ serving as the exchange rate that places the two quantities on a common scale:

Net profit in USD = income in USD − exchange rate × expenditure in EUR

NB in TP units = $TP - \frac{d-b}{a-c} \times FP$

Let us extend this analogy by considering two time periods that represent two subgroups. Suppose a wine importer loses $240,000 in March (subgroup 1) and gains $250,000 in June (subgroup 2), resulting in an overall profit of $10,000. When the same gains and losses are expressed in gold, however, the conclusion reverses. At approximately $92 per gram in March, the $240,000 loss corresponds to about 2,600 g of gold, whereas at $108 per gram in June, the $250,000 gain corresponds to about 2,300 g, resulting in an overall loss of approximately 300 g of gold.

The analogy illustrates the distinction between putting different consequences on a common scale within a subgroup and ensuring that this scale has the same value across subgroups. The exchange rate between US dollars and euros parallels the role of $(d-b)/(a-c)$ in converting false-positive consequences into true-positive units within each subgroup. By contrast, the value of one US dollar in terms of gold differs between March and June; analogously, the utility represented by one unit of NB differs between subgroups when $a-c$ differs. Thus, even after TP and FP consequences have been placed on a common NB scale within each subgroup, subgroup-specific NB values cannot necessarily be aggregated by subgroup size alone without changing the population-utility ranking of strategies.

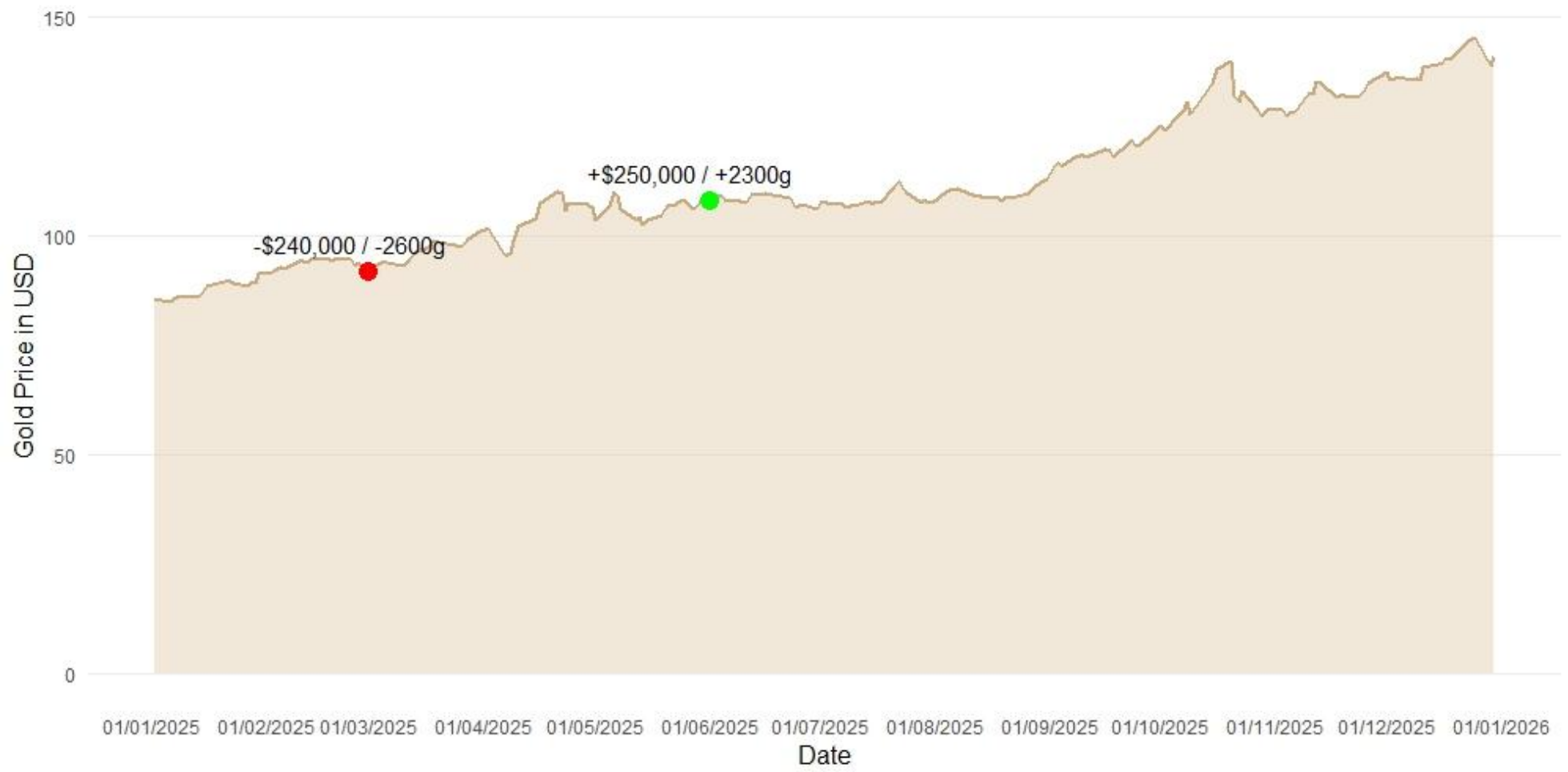

*Figure 3 Gold price in US dollars over time and net profit/loss in US dollars and gold*

## Alternative definitions of NB

Several alternative formulations of NB have been proposed. Although these formulations use different reference outcomes or scaling factors, they do not eliminate the potential inconsistency between population-level NB and population utility when the relevant scaling factor differs across subgroups.

### Net benefit for the untreated

In addition to defining NB from the perspective of individuals classified as positive and therefore recommended for treatment, NB can alternatively be defined from the perspective of individuals classified as negative:

$$NB^{untreated}(t^*) = \frac{1}{N}\left(TN(t^*) - \frac{1-t^*}{t^*}FN(t^*)\right)$$

The utility equation can then be rewritten as:

$$U(t^*) = \frac{d-b}{N}\left(TN(t^*) - \frac{1-t^*}{t^*}FN(t^*)\right) + (a-b)\pi + b$$

This formulation reverses the perspective of conventional NB by interchanging the roles of the outcome and decision categories [10]. The relationship between ΔUtility and ΔNB for the untreated becomes:

$$\Delta U^{untreated} = (d-b)\Delta NB^{untreated}$$

Thus, redefining NB from the perspective of the untreated does not eliminate the potential inconsistency between population-level ΔNB and ΔUtility. Instead, the relevant scaling factor becomes $d-b$, the incremental utility of a true negative relative to a false positive. Consequently, the inconsistency region is determined by the subgroup size ratio and heterogeneity in $d-b$ across subgroups.

### Subgroup net benefit

Benitez-Aurioles et al. [5] proposed an adapted measure termed subgroup net benefit (sNB), which expresses utility using the incremental utility of a true negative relative to a false negative $(d-c)$ as the scaling factor:

$$sNB(t^*) = \frac{U(t^*)}{d-c} = \frac{a-c}{d-c}NB(t^*) - \pi + \frac{d}{d-c}$$

From this definition, the relationship between ΔUtility and ΔsNB can be written as:

$$\Delta U = (d-c)\Delta sNB$$

Thus, sNB does not eliminate the potential inconsistency between population-level ΔUtility and population-level ΔsNB when $d-c$ differs across subgroups. In this formulation, the inconsistency region is determined by the subgroup size ratio and heterogeneity in $d-c$.

# A framework to assess subgroup heterogeneity in DCA

Building on the findings above, conventional population-level DCA may yield conclusions that conflict with population utility when $a-c$ differs across subgroups. We therefore propose a structured framework to assess the robustness of population-level DCA conclusions to subgroup utility heterogeneity.

The framework consists of three steps: Step 1, perform DCA within relevant subgroups; Step 2, determine whether subgroup-specific ΔNB values have the same direction; and Step 3, when ΔNB values have opposite directions, assess whether they fall within the inconsistency region (Figure 4).

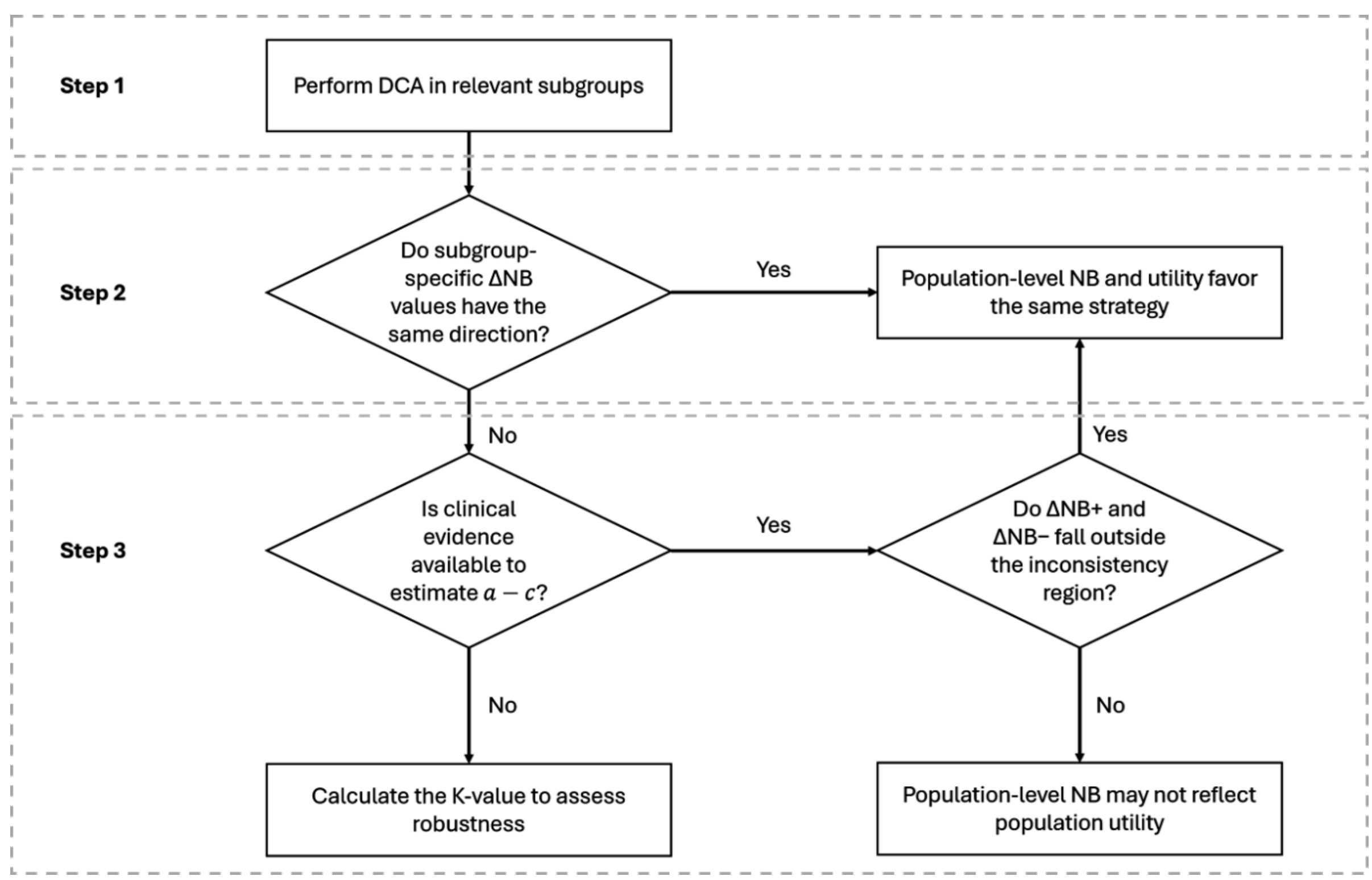

*Figure 4 A framework to explore heterogeneity in DCA*

## Step 1: Perform DCA in subgroups

The size of the inconsistency region is primarily determined by two factors: the relative sizes of the subgroups and differences in $a-c$, the incremental utility of a true positive relative to a false negative. These findings provide a basis for prioritizing subgroup analyses. In practice, priority should be given to clinically relevant subgroups that are sufficiently large to influence population-level conclusions or for which prior clinical evidence suggests meaningful differences in $a-c$. This targeted approach avoids the need to examine subgroups defined by every available patient characteristic. From a fairness perspective, minority subgroups may also warrant evaluation even when their influence on the population-level conclusion is limited.

## Step 2: Check whether △NBs in all subgroups are in the same direction

Once relevant subgroups have been identified, DCA is performed within each subgroup to estimate subgroup-specific ΔNB relative to the comparator strategy. The sign of ΔNB indicates which strategy is favored by NB within each subgroup, while its magnitude indicates the extent of the difference on the NB scale. If ΔNB has the same sign across all subgroups, population-level ΔNB and population utility necessarily favor the same strategy, provided that $a-c>0$ in each subgroup. In this situation, heterogeneity in $a-c$ cannot reverse the population-level ranking of the strategies.

If ΔNB is positive in some subgroups and negative in others, however, the population-level conclusion may depend on how subgroup-specific results are weighted. When subgroup estimates are sufficiently precise, subgroup-specific conclusions may themselves be informative [11]. When subgroup estimates are imprecise, population-level DCA may remain a reasonable starting point, as suggested by Vickers et al. [6]. Nevertheless, opposite signs of subgroup-specific ΔNB provide a warning signal that population-level NB and population utility may favor different strategies, motivating Step 3.

## Step 3: Check the inconsistency conditions

### Define the inconsistency region with clinical evidence

When subgroup-specific ΔNB values have opposite signs, the next step is to assess whether population-level ΔNB and population utility may favor different strategies. For two subgroups, this requires the observed $\Delta NB_1$ and $\Delta NB_2$, the subgroup size ratio $n_1/n_2$, and the relative utility scale $k_1/k_2$, where $k_i = a_i - c_i$.

Evidence on subgroup differences in $a - c$ may be obtained from external clinical studies. Subgroup analyses of clinical trials and stratified or subgroup meta-analyses may provide information on whether intervention effects differ according to characteristics such as age, sex, comorbidity, or disease severity. Although individual subgroup analyses are often exploratory, consistent evidence across studies or appropriately conducted meta-analyses can help inform plausible values of $k_1/k_2$.

The empirical interpretation of $a - c$ differs between diagnostic and prognostic settings. When utility is expressed on the event-risk scale, $a - c$ can be informed by the treatment effect among individuals with the target disease in diagnostic settings. In prognostic settings, under the assumptions described in the Appendix, the relative risk reduction (RRR) provides the corresponding empirical estimate. A similar interpretation has also been suggested in previous methodological studies [5, 12]. A detailed justification is provided in the Appendix.

Using these quantities, the inconsistency region can be determined and the observed pair $(\Delta NB_1, \Delta NB_2)$ located relative to its boundaries. If the point falls outside the inconsistency region, population-level ΔNB and population utility favor the same strategy under the specified $k_1/k_2$. If it falls within the inconsistency region, they favor different strategies, and population-level NB should not be interpreted as a proxy for population utility.

### Calculate the K-value: the critical value of the $k_1/k_2$ ratio

In most practical settings, the exact values of $k_1$ and $k_2$, representing subgroup-specific $a - c$, are unknown. We therefore propose the K-value as a sensitivity measure for assessing the robustness of the population-level NB conclusion to subgroup utility heterogeneity. For an observed pair of $\Delta NB_1$ and $\Delta NB_2$, the K-value is defined as the critical value of the ratio $k_1/k_2$ at which the point lies on the boundary of the inconsistency region. Moving beyond this critical value in the relevant direction implies that population-level NB and population utility favor different strategies.

When $k_1 > k_2$, a pair of $\Delta NB_1$ and $\Delta NB_2$ falls within the inconsistency region when:

$$-\frac{n_1 k_1}{n_2 k_2} < \frac{\Delta NB_2}{\Delta NB_1} < -\frac{n_1}{n_2}$$

The corresponding critical value of $k_1/k_2$ is:

$$\frac{k_1}{k_2} > -\frac{n_2 \Delta NB_2}{n_1 \Delta NB_1} = K$$

Because Step 3 is reached only when the subgroup-specific ΔNB values have opposite signs, $K$ is positive. Thus, when $k_1 > k_2$, population-level NB and population utility favor different strategies if $k_1/k_2 > K$. When $k_1 < k_2$, the calculation of the K-value is analogous, but the direction of the inequality is reversed ($k_1/k_2 < K$). The K-value therefore quantifies how much subgroup-specific $a - c$ must differ before the population-level NB conclusion conflicts with the population-utility conclusion.

## Extend the framework to multiple subgroups

The framework can also be extended to settings with more than two subgroups. Let $i = 1, \ldots, m$ index the subgroups, with subgroup size $n_i$, subgroup-specific $\Delta NB_i$, and $k_i = a_i - c_i$. Population-level ΔNB is proportional to $\sum_i n_i \Delta NB_i$, whereas population-level ΔUtility is proportional to $\sum_i n_i k_i \Delta NB_i$. Thus, population-level ΔNB and ΔUtility favor different strategies when these two sums have opposite signs. As in the two-subgroup setting, such an inconsistency is possible only when subgroup-specific ΔNB values do not all have the same sign, assuming $k_i > 0$ for all subgroups. If all subgroup-specific ΔNB values are positive or all are negative, differences in $k_i$ can change the magnitude of the population-level utility difference but cannot reverse its direction.

When subgroup-specific ΔNB values have different signs, the multiple-subgroup setting can be evaluated by specifying plausible values or ranges for the $k_i$ and examining whether the utility-weighted sum changes sign relative to the sample-size-weighted sum. The same principle therefore applies as in the two-subgroup framework: subgroup size determines each subgroup's contribution to population-level ΔNB, whereas both subgroup size and $a_i - c_i$ determine its contribution to population-level ΔUtility. However, with more than two subgroups, the inconsistency boundary depends jointly on multiple $k_i$ values rather than on a single ratio $k_1/k_2$. Consequently, the two-subgroup K-value does not directly generalize to a single scalar threshold without imposing additional assumptions about the pattern of utility heterogeneity.

# Real-world case study

SCORE2 is a risk prediction algorithm developed for people aged 40–69 years in Europe without prior cardiovascular disease (CVD) or diabetes to estimate their 10-year risk of a first cardiovascular event [13]. In a hypothetical clinical decision scenario, we evaluated SCORE2 at a 3% threshold probability, whereby LDL-lowering therapy would be initiated in individuals with an estimated 10-year CVD risk above this threshold. A systematic review reported a moderate effect of LDL-lowering therapy on major vascular events among people without known vascular disease (RRR=25% per 1.0 mmol/L reduction in LDL cholesterol, corresponding to a rate ratio of 0.75) [14].

## Step 1: Perform DCA in subgroups

An external validation study of SCORE2 reported that, at a 3% threshold and assuming equal proportions of men and women, the population-level NB of SCORE2 was 0.04560 [15]. The corresponding NB of "treat all" was 0.04468, yielding a population-level ΔNB of 0.00092 in favor of SCORE2. Subgroup-specific NB for SCORE2 was 0.05885 in men and 0.03235 in women (Table 3).

## Step 2: Check whether △NBs in all subgroups are in the same direction

The subgroup-specific ΔNB values had opposite signs. In men, the NB of SCORE2 (0.05885) was slightly lower than that of "treat all" (0.06045), yielding $\Delta NB_{\text{Men}} = -0.00160$; thus, NB favored "treat all". In women, the NB of SCORE2 (0.03235) exceeded that of "treat all" (0.02891), yielding $\Delta NB_{\text{Women}} = 0.00345$; thus, NB favored SCORE2 (Table 3) [15]. Because the subgroup-specific ΔNB values had opposite signs, we proceeded to Step 3 to assess whether population-level NB and population utility could favor different strategies.

## Step 3: Check the inconsistency conditions

### Define the inconsistency region using clinical evidence

We first assessed the inconsistency region using external evidence on subgroup-specific $a - c$ (Figure 5). Here, $a - c$ represents the incremental utility of initiating LDL-lowering therapy versus not initiating treatment among individuals who would experience the event in the absence of treatment. Because SCORE2 is a prognostic prediction model, we used the RRR associated with LDL-lowering therapy as the empirical estimate of $a - c$, rather than the absolute risk reduction. A detailed justification is provided in the Appendix.

A sex-specific meta-analysis reported RRRs of 0.28 (rate ratio=0.72) in men and 0.15 (rate ratio=0.85) in women [14], corresponding to an $a - c$ ratio of $0.28/0.15 = 1.867$ (Table 3). Under these treatment-effect estimates, the observed point $(\Delta NB_{\text{Men}}, \Delta NB_{\text{Women}})$ lies outside the inconsistency region (the gray area between the black and blue dashed lines in Figure 5). Therefore, population-level NB and population utility favor the same strategy, SCORE2.

To illustrate how greater treatment-effect heterogeneity could alter this conclusion, suppose hypothetically that the rate ratio in women was 0.88 (RRR=0.12) rather than 0.85. The corresponding $a - c$ ratio would then be $0.28/0.12 = 2.333$, placing the observed pair $(\Delta NB_{\text{Men}}, \Delta NB_{\text{Women}})$ within the inconsistency region (between the black and green dashed lines in Figure 5). In this scenario, population-level NB would favor SCORE2, whereas population utility would favor "treat all".

### Calculate the K-value to assess robustness

If subgroup-specific $a - c$ values are unknown, the K-value provides an alternative robustness assessment. Assuming equal subgroup sizes, the K-value in this example is

$$K = -\frac{n_{\text{Women}}\Delta NB_{\text{Women}}}{n_{\text{Men}}\Delta NB_{\text{Men}}} = -\frac{0.00345}{-0.00160} = 2.156$$

Thus, because $a - c$ is assumed to be larger in men than in women, population-level NB and population utility would favor different strategies only if the ratio $(a - c)_{\text{Men}}/(a - c)_{\text{Women}}$ exceeded 2.156. The evidence-based ratio of 1.867 therefore remains below the K-value, whereas the hypothetical ratio of 2.333 exceeds it.

## Epilogue to the case study

This case study is intended as a methodological illustration rather than a clinical recommendation regarding the use of SCORE2 or LDL-lowering therapy. The 3% threshold was selected specifically to illustrate a setting in which subgroup-specific ΔNB values have opposite signs. Therefore, the resulting comparison between SCORE2 and "treat all" should not be interpreted as evidence for either strategy in clinical practice.

In this case, the negative ΔNB for SCORE2 relative to "treat all" in men may be attributable to model miscalibration. Previous work has shown that, with adequate calibration, the ΔNB of a prediction model relative to the relevant default strategy should be non-negative [16]. Adequate recalibration would therefore be expected to eliminate the negative ΔNB and, consequently, the conflicting signs of subgroup-specific ΔNB observed in Step 2. If ΔNB is then consistently non-negative across subgroups, the population-level NB conclusion would no longer be vulnerable to the inconsistency examined in our framework, and use of the model across the population may be supported by DCA. However, decision-makers appraising an existing model may not have access to the data required for recalibration. Thus, our framework remains relevant when model updating is not feasible.

Finally, when a prediction model is used to guide treatment initiation, the relevant predicted risk should ideally represent the risk in the absence of treatment. If treatment is initiated during follow-up and is not appropriately accounted for during model development, predicted risks may reflect a mixture of treated and untreated outcomes, complicating their interpretation for treatment decisions [17-19]. This issue is relevant to SCORE2 because preventive treatment initiated during follow-up was not accounted for during model development [13].

*Table 3 Net benefits and treatment effects in men and women*

| | **Model Performance** | | | **Treatment Effect** | |
|---|---|---|---|---|---|
| | **NB(SCORE 2)** | **NB(Tx All)** | **ΔNB** | **RR for MVE** | **RRR for MVE (a-c estimate)** |
| **Men** | 0.05885 | 0.06045 | -0.00160 | 0.72 | 0.28 |
| **Women** | 0.03235 | 0.02891 | 0.00345 | 0.85 | 0.15 |
| **All** | 0.04560 | 0.04468 | 0.00092 | | |

*Note: Net benefit data were extracted from Extended Data Figure 3 in [15]; treatment-effect and event-rate data were extracted from Figure 1 in [14]. RR=rate ratio; RRR=relative risk reduction; MVE=major vascular event.*

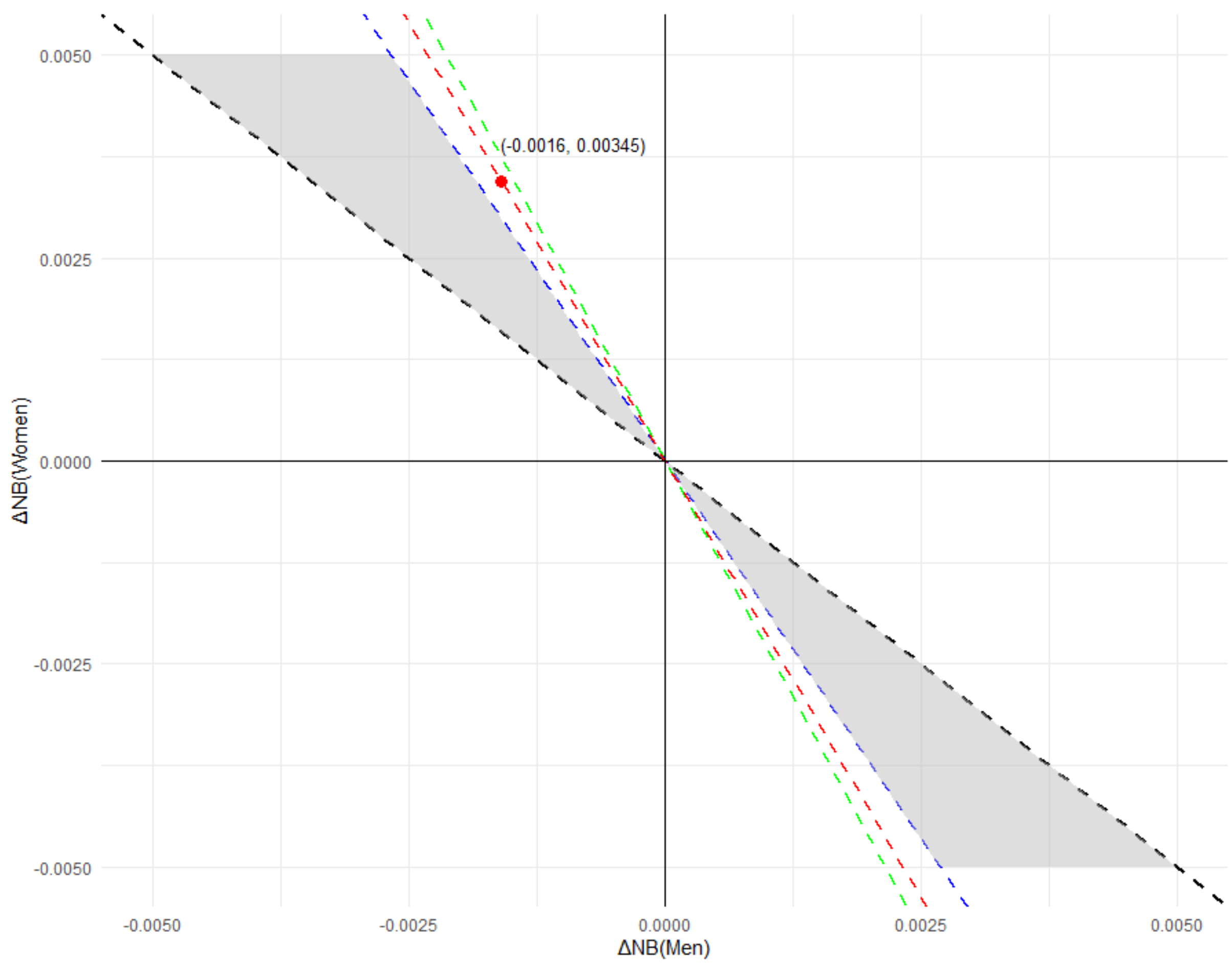

*Figure 5 Inconsistency region for SCORE2 by sex*

*Note: The black dashed line is $y = -x$, corresponding to equal subgroup sizes ($n_{men} = n_{women}$). The blue dashed line is $y = -1.867x$, corresponding to the evidence-based ratio $n_{men}k_{men}/n_{women}k_{women} = 1.867$, based on the sex-specific relative risk reductions for major vascular events. The green dashed line is $y = -2.333x$, corresponding to the hypothetical ratio $n_{men}k_{men}/n_{women}k_{women} = 2.333$. The observed point ($\Delta NB_{men}, \Delta NB_{women}$) lies on the red dashed line, $y = -2.156x$, where $K = -(n_{women}\Delta NB_{women})/(n_{men}\Delta NB_{men}) = 2.156$ is the critical value for inconsistency.*

# Discussion

In this study, we demonstrated that interpreting population-level NB as a proxy for population utility can lead to incorrect conclusions in the presence of subgroup utility heterogeneity. When subgroup-specific ΔNB values have opposite signs, population-level NB may favor prediction model use while population utility favors the comparator strategy, or vice versa. Such inconsistencies arise when $a - c$, the incremental utility of a true positive relative to a false negative, differs across subgroups, and become more likely when these differences are large and subgroup sizes are similar.

The central insight of this study is that quantities expressed in different units should not be directly aggregated. Because NB expresses utility in units of $a - c$, subgroup-specific NB values are expressed on different implicit utility scales when $a - c$ differs across subgroups. Population-level NB weights these subgroup-specific values according to subgroup size, whereas population utility additionally reflects subgroup-specific $a - c$. Consequently, averaging subgroup-specific NB values can produce a population-level ranking of strategies that differs from the ranking based on population utility. From a statistical perspective, this discrepancy can be viewed as a form of non-collapsibility: NB is valid as a utility-based comparison within each subgroup, but population-level NB need not remain proportional to population utility after pooling subgroups with different $a - c$ values.

A second methodological contribution of this study is linking the abstract utility parameter $a - c$ to empirical treatment-effect evidence. The empirical interpretation differs between diagnostic and prognostic settings. For diagnostic models, $a - c$ can be informed directly by the treatment effect among individuals with the target disease. For prognostic models, under the assumptions described in the Appendix, the RRR provides the corresponding empirical estimate of $a - c$. This interpretation is consistent with previous methodological work proposing the integration of RRRs into the DCA framework [5, 12]. This link allows subgroup utility heterogeneity to be informed by external clinical evidence rather than requiring direct elicitation of all underlying utility values.

Previous guidance has recommended conducting subgroup DCA only when subgroup-specific conclusions are supported by compelling evidence, partly because subgroup analyses often involve limited data [6]. This parallels the conventional emphasis on population-average treatment effects in randomized trials, but an important distinction arises in DCA. Treatment-effect heterogeneity in a randomized trial does not, by itself, invalidate the average treatment effect. In DCA, however, heterogeneity in subgroup-specific $a - c$ can cause population-level NB and population utility to favor different strategies. Subgroup evaluation may therefore be important not only for identifying heterogeneity in prediction model usefulness or addressing fairness concerns, but also for assessing whether the population-level NB conclusion is robust.

Our framework provides a practical approach to this assessment by first using subgroup-specific ΔNB to identify situations in which inconsistency is possible and then incorporating plausible heterogeneity in $a - c$ from external evidence. When subgroup-specific $a - c$ values are unavailable, the K-value provides a sensitivity measure by quantifying the critical $a - c$ ratio at which population-level NB and population utility begin to favor different strategies. Analogous to the E-value [20], which quantifies the impact of unmeasured confounding, the K-value provides an intuitive measure of the robustness of population-level DCA conclusions to potential subgroup utility heterogeneity.

This study has several limitations. First, we take a model evaluation perspective, focusing specifically on prediction model usefulness as assessed by DCA rather than the broader process of model adoption, which ideally should also consider evidence of clinical impact and implementation factors, although prediction models may in practice enter clinical use without formal evaluation of their clinical usefulness. Consistent with this perspective, we did not include recalibration as a step in our framework. Negative subgroup-specific ΔNB may arise from model miscalibration, and adequate recalibration can eliminate such negative ΔNB and thereby resolve conflicting signs across subgroups [16]. However, decision-makers appraising an existing model may not have access to the data required for model updating. Our framework is therefore intended to assess the robustness of the available DCA evidence when recalibration is not feasible. Second, our mathematical results use the same conventional DCA framework for diagnostic and prognostic models, consistent with standard DCA definitions [2, 21]. Although the empirical interpretation of $a - c$ differs between these settings, as discussed above and detailed in the Appendix, generalized DCA can explicitly account for differences in event probabilities under treatment [12]. Third, our findings do not directly apply to prediction models that estimate individualized differences in outcome risk under alternative treatment strategies [7], rather than absolute outcome risks. Extension of the framework to such settings requires further investigation. Finally, although our analyses focused primarily on comparisons with the default strategies “treat all” and “treat none”, the same principles extend to comparisons between competing prediction models or with current practice.

# Conclusion

This study demonstrates that population-level NB and population utility can favor different strategies when subgroups with different $a - c$ values are pooled. This finding highlights the importance of considering subgroup utility heterogeneity when interpreting population-level NB as a proxy for population utility. Our framework provides a practical approach to identifying situations in which such heterogeneity may lead to inconsistent conclusions and to assessing the robustness of population-level DCA conclusions to potential subgroup utility heterogeneity.

# Acknowledgments

## Use of AI tool

ChatGPT (OpenAI) was used solely to assist with language refinement and to improve consistency in writing style across the manuscript. The authors retain full responsibility for the scientific content, interpretation, and conclusions of the manuscript.

## Funding

This research is supported by the Netherlands Organization for Scientific Research (NWO) under the VIDI grant 09150172310023.

# Appendix

## Proofs

### Proof of the relationship between ΔUtility and ΔNB

If we compare the model with "treat all":

$$\Delta U = U(t^*) - U(Tx\ all) = (a-c)NB(t^*) + (c-d)\pi + d - (a\pi + (1-\pi)\mathrm{b})$$
$$= (a-c)NB(t^*) - \left((a-c)\pi - (\mathrm{d}-\mathrm{b})(1-\pi)\right)$$
$$= (a-c)NB(t^*) - (a-c)\left(\pi - \frac{d-b}{a-c}(1-\pi)\right)$$
$$= (a-c)NB(t^*) - (a-c)\left(\pi - \frac{t^*}{1-t^*}(1-\pi)\right) = (a-c)NB(t^*) - (a-c)NB(Tx\ all)$$
$$= (a-c)\Delta NB$$

If we compare the model with "treat none":

$$\Delta U = U(t^*) - U(Tx\ none) = (a-c)NB(t^*) + (c-d)\pi + d - (c\pi + (1-\pi)\mathrm{d}) = (a-c)NB(t^*) - 0$$
$$= (a-c)NB(t^*) - (a-c)NB(Tx\ none) = (a-c)\Delta NB$$

If we compare "treat all" vs "treat none":

$$\Delta U = U(Tx\ all) - U(Tx\ none) = (a\pi + (1-\pi)\mathrm{b}) - (c\pi + (1-\pi)\mathrm{d}) = (a-c)\pi - (d-b)(1-\pi)$$
$$= (a-c)\left(\pi - \frac{d-b}{a-c}(1-\pi)\right) = (a-c)NB(Tx\ all) - 0$$
$$= (a-c)NB(Tx\ all) - (a-c)NB(Tx\ none) = (a-c)\Delta NB$$

If we compare model A with model B, the relationship can be derived using a common default strategy, such as "treat none", as a reference:

$$\Delta U = U(Model\ A) - U(Model\ B) = (U(Model\ A) - U(Tx\ none)) - (U(Model\ B) - U(Tx\ none))$$
$$= (a-c)(NB(Model\ A) - NB(Tx\ none)) - (a-c)(NB(Model\ B) - NB(Tx\ none))$$
$$= (a-c)(NB(Model\ A) - NB(Model\ B)) = (a-c)\Delta NB$$

### Proof that the sample-size-weighted subgroup ΔNB equals population-level ΔNB

Assuming a common threshold probability $t^*$ across subgroups, population-level ΔNB is the sample-size-weighted (with $N = n_1 + n_2$) average of subgroup-specific ΔNB:

$$\Delta NB_{1+2} = \frac{1}{N}(n_1\Delta NB_1 + n_2\Delta NB_2) = \frac{1}{N}(n_1(NB_1(t^*) - NB_1(Tx\ all)) + n_2(NB_2(t^*) - NB_2(Tx\ all)))$$
$$= \frac{1}{N}\left(n_1\left(\frac{TP_1(t^*)}{n_1} - \frac{t^*}{1-t^*}\frac{FP_1(t^*)}{n_1}\right) - n_1\left(\pi_1 - \frac{t^*}{1-t^*}(1-\pi_1)\right)\right.$$
$$\left. + n_2\left(\frac{TP_2(t^*)}{n_2} - \frac{t^*}{1-t^*}\frac{FP_2(t^*)}{n_2}\right) - n_2\left(\pi_2 - \frac{t^*}{1-t^*}(1-\pi_2)\right)\right)$$
$$= \frac{1}{N}\left(TP_1(t^*) + TP_2(t^*) - \frac{t^*}{1-t^*}\left(FP_1(t^*) + FP_2(t^*)\right)\right)$$
$$- \frac{1}{N}\left(n_1\pi_1 + n_2\pi_2 - \frac{t^*}{1-t^*}(n_1(1-\pi_1) + n_2(1-\pi_2))\right)$$
$$= \frac{1}{N}\left(TP - \frac{t^*}{1-t^*}FP\right) - \left(\frac{n_1\pi_1 + n_2\pi_2}{N} - \frac{t^*}{1-t^*}\frac{\left(n_1 + n_2 - (n_1\pi_1 + n_2\pi_2)\right)}{N}\right)$$
$$= \frac{1}{N}\left(TP - \frac{t^*}{1-t^*}FP\right) - \left(\pi - \frac{t^*}{1-t^*}(1-\pi)\right) = NB(t^*) - NB(Tx\ all) = \Delta NB$$

## Proof of the inconsistency region

Let $\Delta NB_1 = x$, $\Delta NB_2 = y$, $a_1 - c_1 = k_1$, and $a_2 - c_2 = k_2$.
The function $f(x, y)$ can be expressed as a quadratic form with symmetric matrix $Q$:

$$f(x, y) = ({n_1}^2 k_1 x^2 + n_1 n_2 (k_1 + k_2) xy + {n_2}^2 k_2 y^2) = [x \quad y] \begin{bmatrix} {n_1}^2 k_1 & \frac{1}{2} n_1 n_2 (k_1 + k_2) \\ \frac{1}{2} n_1 n_2 (k_1 + k_2) & {n_2}^2 k_2 \end{bmatrix} \begin{bmatrix} x \\ y \end{bmatrix}$$

$$Q = \begin{bmatrix} {n_1}^2 k_1 & \frac{1}{2} n_1 n_2 (k_1 + k_2) \\ \frac{1}{2} n_1 n_2 (k_1 + k_2) & {n_2}^2 k_2 \end{bmatrix}$$

We assess the definiteness of $Q$ using Sylvester's criterion:

$$\det(Q_1) = n_1^2 k_1 > 0 \ (k_1 > 0)$$

$$\det(Q_2) = \left({n_1}^2 k_1 {n_2}^2 k_2 - \frac{1}{4} {n_1}^2 {n_2}^2 (k_1 + k_2)^2\right) = {n_1}^2 {n_2}^2 \left(k_1 k_2 - \frac{1}{4}(k_1 + k_2)^2\right) = -\frac{1}{4} {n_1}^2 {n_2}^2 (k_1 - k_2)^2$$

$$\det(Q_2) = -\frac{1}{4} n_1^2 n_2^2 (k_1 - k_2)^2 < 0 \ (k_1 \neq k_2)$$

Therefore, when $k_1 \neq k_2$, $f(x, y)$ is an indefinite quadratic form, implying that there exist combinations of $(x, y)$ for which $f(x, y) < 0$. These combinations define the inconsistency region.

The inconsistency region is bounded by $f(x, y) = 0$, which consists of two straight lines intersecting at the origin. Let $y = sx$. Substitution gives:

$$f(x, y) = ({n_1}^2 k_1 x^2 + n_1 n_2 (k_1 + k_2) xy + {n_2}^2 k_2 y^2) = x^2 ({n_1}^2 k_1 + n_1 n_2 (k_1 + k_2) s + {n_2}^2 k_2 s^2) = 0$$

$${n_1}^2 k_1 + n_1 n_2 (k_1 + k_2) s + {n_2}^2 k_2 s^2 = 0$$

$$s = \frac{-n_1 n_2 (k_1 + k_2) \pm \sqrt{{n_1}^2 {n_2}^2 (k_1 + k_2)^2 - 4{n_1}^2 {n_2}^2 k_1 k_2}}{2{n_2}^2 k_2} = \frac{-n_1 n_2 (k_1 + k_2) \pm \sqrt{{n_1}^2 {n_2}^2 (k_1 - k_2)^2}}{2{n_2}^2 k_2}$$
$$= \frac{-n_1 n_2 (k_1 + k_2) \pm n_1 n_2 (k_1 - k_2)}{2{n_2}^2 k_2}$$

$$s_1 = \frac{-n_1 n_2 (k_1 + k_2) + n_1 n_2 (k_1 - k_2)}{2{n_2}^2 k_2} = \frac{-2 n_1 n_2 k_2}{2{n_2}^2 k_2} = -\frac{n_1}{n_2}$$
$$s_2 = \frac{-n_1 n_2 (k_1 + k_2) - n_1 n_2 (k_1 - k_2)}{2{n_2}^2 k_2} = \frac{-2 n_1 n_2 k_1}{2{n_2}^2 k_2} = -\frac{n_1 k_1}{n_2 k_2}$$

Alternatively, the two boundary lines can be obtained directly by factorizing $f(x, y)$:

$$f(x, y) = (n_1 x + n_2 y)(n_1 k_1 x + n_2 k_2 y)$$

Thus, $f(x, y) = 0$ when either $n_1 x + n_2 y = 0$ or $n_1 k_1 x + n_2 k_2 y = 0$, which gives:

$$y = -\frac{n_1}{n_2} x \text{ or } y = -\frac{n_1 k_1}{n_2 k_2} x$$

## Estimating $a - c$ from external clinical evidence

Our framework requires specification of $a - c$, the incremental utility of a true positive relative to a false negative. When the prediction model is used to guide treatment, $a - c$ can be interpreted as the incremental utility of treatment versus no treatment among individuals who would experience the outcome in the absence of treatment. This quantity may be informed by external clinical evidence, although its empirical interpretation differs between diagnostic and prognostic settings.

### Diagnostic prediction models

For diagnostic prediction models, the empirical interpretation of $a - c$ is relatively straightforward. A true positive is an individual with the target disease who is correctly identified and treated, whereas a false negative is an individual with the target disease who is not treated. Thus, $a - c$ represents the incremental utility of treatment versus no treatment among individuals with the target disease.

Randomized trials evaluating treatment are typically conducted among individuals with confirmed disease. The treatment effect estimated in such trials therefore corresponds directly to the average treatment effect among individuals with the target disease, $E(\text{treatment effect} \mid D = 1)$, where $D$ denotes true disease status. Here, treatment effect is understood on the same outcome or utility scale used to define $a - c$; for example, when expressed on the event-risk scale, it corresponds to the absolute risk reduction.

Because the trial population corresponds to the population for whom $a - c$ is defined, no additional rescaling is required. The treatment effect estimated from the randomized trial can therefore be used directly as an empirical estimate of $a - c$.

### Prognostic prediction models

For prognostic prediction models, the situation is different because individuals who would experience the future outcome in the absence of treatment cannot be identified at the time of trial enrollment. Randomized trials therefore enroll individuals before the outcome occurs and estimate the average treatment effect across a population comprising both individuals who would and would not experience the outcome if untreated.

Randomized trials may quantify the average treatment effect on the event-risk scale using the absolute risk reduction:

$$P(Y = 1 \mid \text{untreated}) - P(Y = 1 \mid \text{treated})$$

which represents the average reduction in event risk across the entire trial population.

Under our simplifying decision-analytic framework, treatment is assumed to benefit only individuals who would experience the event if untreated and to have no effect on individuals who would remain event-free without treatment. The average treatment effect on the event-risk scale can therefore be decomposed as:

$$P(Y = 1 \mid \text{untreated}) - P(Y = 1 \mid \text{treated}) = E(\text{treatment effect} \mid Y = 1, \text{untreated}) \times P(Y = 1 \mid \text{untreated}) + 0 \times P(Y = 0 \mid \text{untreated})$$

Here, $E(\text{treatment effect} \mid Y = 1, \text{untreated})$ represents the average treatment effect among individuals who would experience the event if untreated, which corresponds to $a - c$. Rearranging the expression above gives

$$a - c = E(\text{treatment effect} \mid Y = 1, \text{untreated}) = \frac{P(Y = 1 \mid \text{untreated}) - P(Y = 1 \mid \text{treated})}{P(Y = 1 \mid \text{untreated})}$$

which is equivalent to the relative risk reduction (RRR). Under the simplifying assumptions above, the RRR from randomized clinical trials can therefore be used as an empirical estimate of $a - c$ for prognostic prediction models.

## Scenario analysis

To explore the inconsistency region under a range of practically relevant settings, we conducted a scenario analysis using different combinations of subgroup characteristics, threshold probabilities, and model performance. Because the inconsistency region is determined by the subgroup size ratio $n_1/n_2$ and the ratio of subgroup-specific $a - c$ values, $(a_1 - c_1)/(a_2 - c_2)$, we considered four main scenarios defined by $n_1/n_2$=1 or 10 and $(a_1 - c_1)/(a_2 - c_2)$=2 or 4. Within each scenario, we varied the threshold probability (0.2, 0.33, 0.5, 0.67 and 0.8; corresponding threshold odds 0.25, 0.5, 1, 2 and 4), sensitivity (0.5–0.95), and specificity (0.5–0.95). We assumed the same threshold probability and outcome prevalence across subgroups, considering prevalence values of $\pi = 0.5$ and $\pi = 0.3$, with $d = 100$ and $c = 0$. ΔNB was calculated relative to the best default strategy: "treat all" when the threshold probability was below the outcome prevalence and "treat none" otherwise [2].

The scenario analysis shows that larger between-subgroup differences in $a - c$ increase the likelihood that population-level NB and population utility favor different strategies. This inconsistency occurs more frequently at threshold probabilities farther from the outcome prevalence.

The eight panels on the left in Figure S 1 illustrate the inconsistency region as the gray area between the two red dotted boundary lines, which are determined by $n_1/n_2$ and $(a_1 - c_1)/(a_2 - c_2)$. The pairs of subgroup-specific $(\Delta NB_1, \Delta NB_2)$ generated by different combinations of sensitivity and specificity at the five threshold probabilities are shown as colored dots. As the subgroup size ratio $n_1/n_2$ increases, the inconsistency region becomes narrower, reflecting the increasing influence of the larger subgroup on the population-level ΔNB. Conversely, for a fixed $n_1/n_2$, a larger difference between subgroup-specific $a - c$ values, represented by an increase in $(a_1 - c_1)/(a_2 - c_2)$, results in a larger inconsistency region.

The eight panels on the right in Figure S 1 show the relationship between population-level ΔNB and ΔUtility. The upper-left and lower-right quadrants represent the inconsistency region, in which ΔNB and ΔUtility have opposite signs and therefore favor different strategies. For each threshold probability, the combinations of sensitivity and specificity generate a parallelogram-shaped region of possible $(\Delta NB, \Delta Utility)$ values. Whether a particular combination falls within the inconsistency region therefore depends on the threshold probability and the corresponding model performance. When $a_1 - c_1 = a_2 - c_2$, population-level ΔUtility is proportional to population-level ΔNB, so all pairs lie on a straight line through the origin with positive slope and no inconsistency is possible. As the ratio $(a_1 - c_1)/(a_2 - c_2)$ increases, the parallelogram-shaped regions become wider, increasing their overlap with the inconsistency region, particularly when subgroup sizes are equal.

Similar results are observed in Figure S 2 for $\pi$=0.3.

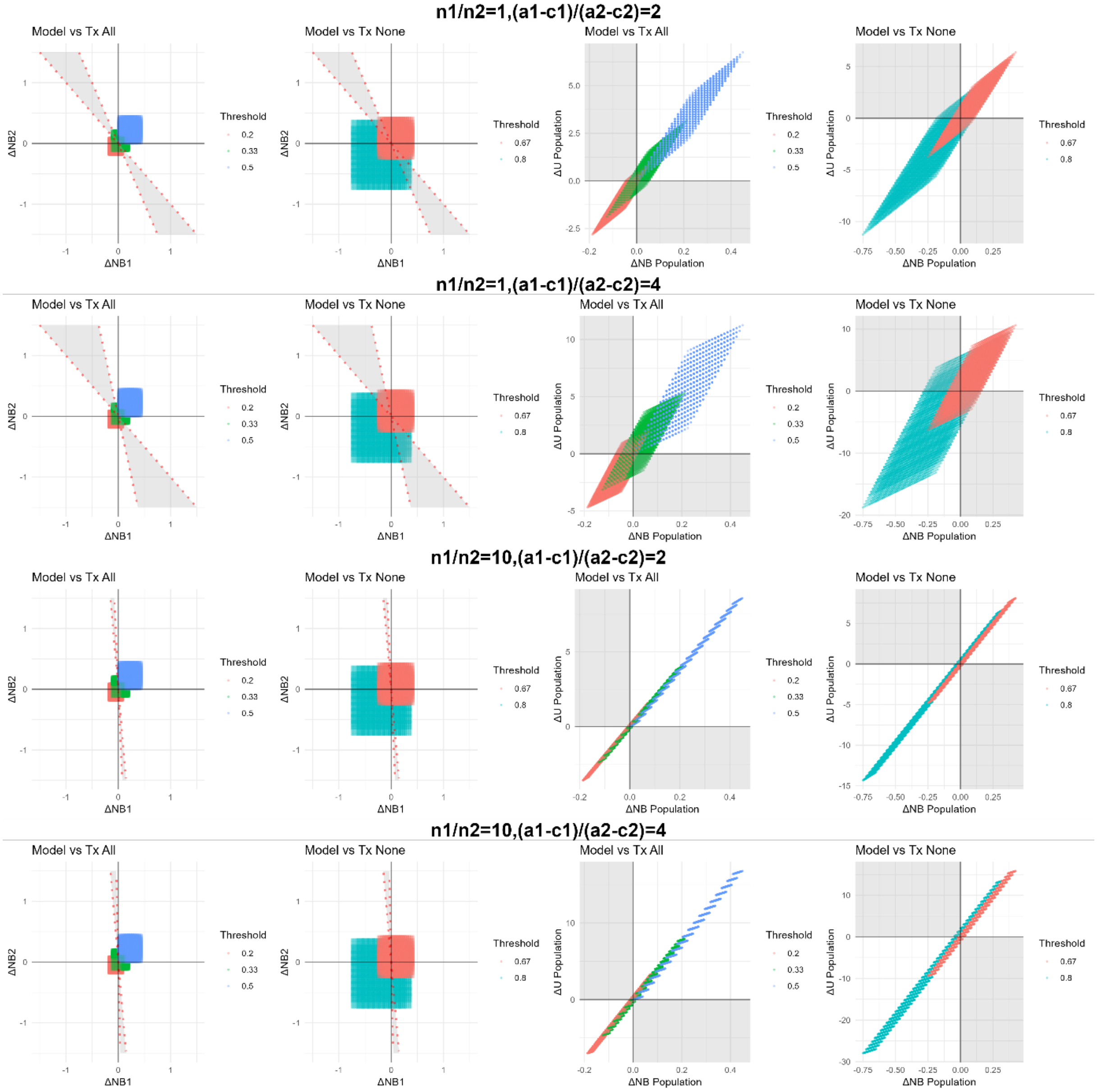

*Figure S 1 Scenario analysis ($\pi$=0.5)*

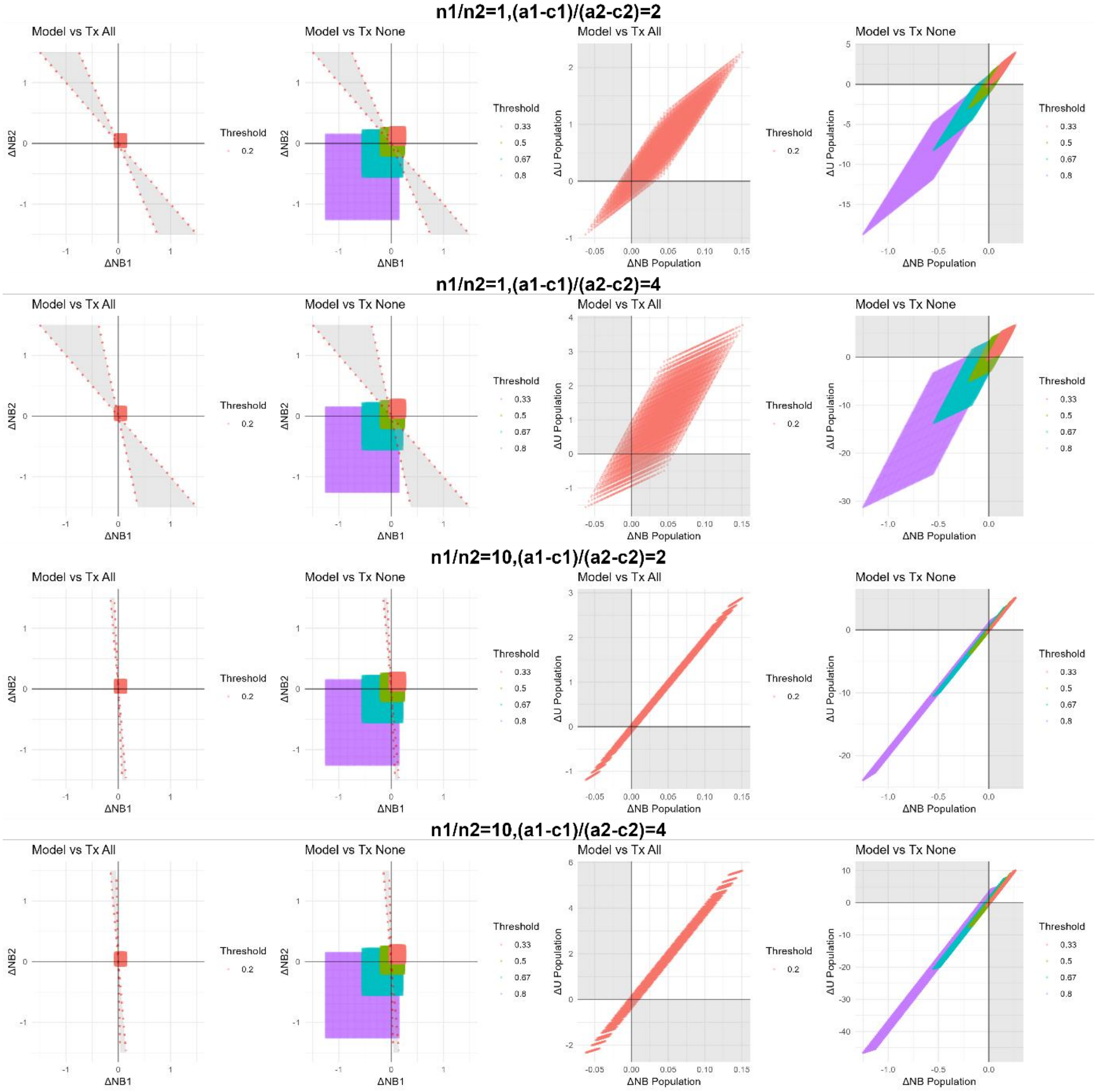

*Figure S 2 Scenario analysis ($\pi$=0.3)*